\documentclass[12pt]{article}

\newcommand{\m}{\mathrm}
\newcommand{\be}{\begin{equation}}
\newcommand{\ee}{\end{equation}}
\newcommand{\ba}{\begin{eqnarray}}
\newcommand{\ea}{\end{eqnarray}}

\usepackage[nosort]{cite}
\usepackage{graphicx}
\usepackage[font={scriptsize,singlespacing}]{caption}
\usepackage{epstopdf}
\usepackage{amsmath}
\usepackage{amssymb}
\usepackage{subfigure}
\usepackage{hyperref}
\usepackage{url}
\usepackage{xcolor}
\usepackage{color}

\definecolor{purple(munsell)}{rgb}{0.62, 0.0, 0.77}
\definecolor{palatinateblue}{rgb}{0.15, 0.23, 0.89}
\definecolor{royalblue(web)}{rgb}{0.25, 0.41, 0.88}
\hypersetup{linktoc=all,
    colorlinks, linkcolor={purple(munsell)},
    citecolor={palatinateblue}, urlcolor={palatinateblue}
}

\def\sideremark#1{\ifvmode\leavevmode\fi\vadjust{\vbox to0pt{\vss
 \hbox to 0pt{\hskip\hsize\hskip1em
 \vbox{\hsize2cm\tiny\raggedright\pretolerance10000
 \noindent #1\hfill}\hss}\vbox to8pt{\vfil}\vss}}}%
\begin{document}
\thispagestyle{empty}
\begin{center}

\null \vskip-1truecm \vskip2truecm

{\Large{\bf \textsf{When Is Holography Consistent?}}}

{\large{\bf \textsf{}}}

{\large{\bf \textsf{}}}

\vskip1truecm

{\large \textbf{\textsf{Brett McInnes}}}

\textsf{  National
  University of Singapore}\\

\texttt{Email: matmcinn@nus.edu.sg}
\vskip0.3truecm
and
\vskip0.3truecm
{\large\textbf{\textsf{Yen Chin Ong}}}

\textsf{Nordita, \\ KTH Royal Institute of Technology and Stockholm University,\\  Roslagstullsbacken 23,
SE-106 91 Stockholm, Sweden}\\
\texttt{Email:  yenchin.ong@nordita.org}

\vskip1truecm

\end{center}
\vskip1truecm \centerline{\textsf{ABSTRACT}} \baselineskip=15pt
\medskip

Holographic duality relates two radically different kinds of theory: one with gravity, one without. The very existence of such an equivalence imposes strong consistency conditions which are, in the nature of the case, hard to satisfy. Recently a particularly deep condition of this kind, relating the minimum of a probe brane action to a gravitational bulk action (in a Euclidean formulation), has been recognised; and the question arises as to the circumstances under which it, and its Lorentzian counterpart, are satisfied. We discuss the fact that there are physically interesting situations in which one or both versions might, in principle, \emph{not} be satisfied. These arise in two distinct circumstances: first, when the bulk is not an Einstein manifold, and, second, in the presence of angular momentum. Focusing on the application of holography to the quark-gluon plasma (of the various forms arising in the early Universe and in heavy-ion collisions), we find that these potential violations never actually occur. This suggests that the consistency condition is a ``law of physics'' expressing a particular aspect of holography.

\newpage
\addtocounter{section}{1}
\section* {\large{\textsf{1. A Holographic Consistency Condition}}}
The essence of holographic duality \cite{kn:veron}, and the reason for its usefulness \cite{kn:makoto}, is that it equates two systems with apparently very different natures and degrees of freedom. Intuitively, it must in some sense be ``difficult'' to arrange this; more formally, there must be a whole network of \emph{consistency conditions} between bulk and boundary which make it possible.

Recently it has been shown that a deep study \cite{kn:ferrari1,kn:ferrari2} of \emph{probe branes} in the (Euclidean) bulk can yield a new perspective on holography. In particular, one is led in this way to a fundamental consistency condition \cite{kn:ferrari3},
\begin{equation}\label{A}
S_g^*\;=\;{N\over \gamma}S_b^*,
\end{equation}
where $S_g^*$ is the (on-shell) gravitational action in the bulk, $N$ is the number of colours in the boundary theory, $\gamma$ is the scaling exponent for the free energy of the boundary theory, and $S_b^*$ is the probe brane action.

As is emphasised in \cite{kn:ferrari3}, it is very remarkable that a relation between such different objects as $S_g^*$ and $S_b^*$ can hold at all, and certainly it is not to be expected that it will \emph{always} be valid. The problem is to understand the circumstances in which it is valid; in particular, one needs to show that \emph{no reasonably realistic physical system is in conflict with it}.

A rather simple example of a bulk geometry which does \emph{not} satisfy this relation is as follows. Following \cite{kn:ferrari3}, we focus on the Euclidean domain; then it is possible to prove that equation (\ref{A}) will hold if the ($d+1$)-dimensional bulk is an Einstein manifold (like anti-de Sitter space itself) and the following condition is satisfied: for each hypersurface $\Sigma$ in the bulk homologous to the boundary, the area\footnote{If these areas and volumes are infinite (because the boundary may not be compact) then we arbitrarily choose a compact domain in $\Sigma$ and interpret area and volume to refer to that domain.} $A(\Sigma)$ and the volume $V(M_{\Sigma})$ enclosed by $\Sigma$ should satisfy the ``isoperimetric inequality''
\begin{equation}\label{B}
\mathfrak{S^{\m{E}}}\;\equiv\;A(\Sigma)\;- \;{d \over L}V(M_{\Sigma}) \;\geqslant \;0,
\end{equation}
where, here and henceforth, $L$ denotes the asymptotic curvature scale, and the superscript ``E'' denotes a Euclidean quantity. This condition is certainly satisfied by all surfaces in Euclidean $AdS_{d+1}$ homologous to the boundary, and one can show that it is satisfied by the relevant surfaces in the Euclidean AdS-Schwarzschild geometry when the (Lorentzian) event horizon has a spherical topology.

As is well known \cite{kn:lemmo,9607071,9609065,9808032}, however, the event horizon of an asymptotically AdS black hole need not have spherical topology and geometry; it can, for example, be modelled on a manifold of constant \emph{negative} curvature. Let $X^{-1}_2$ be a compact two-dimensional manifold of curvature $- 1$ (a Riemann surface), let d$\Omega^2[X_2^{-1}]$ be its (dimensionless) metric, and let $V[X^{-1}_2]$ be the corresponding volume. A (Euclidean) four-dimensional asymptotically AdS dyonic Reissner-Nordstr\"om black hole metric of this kind, with mass, electric charge, and magnetic charge parameters $M$, $Q$, and $P$, takes the form\footnote{Notice that the coefficient of $Q^2$ is the opposite of that of $P^2$ in the Euclidean case: see below.}
\begin{eqnarray}\label{C}
g^E(\m{AdSdyRN^{-1}_{4})} &=&  \Bigg[{r^2\over L^2}\;-\;1\;-\;{8\pi M\over V[X^{-1}_2] r}\;+\;{4\pi \left (-\,Q^2 + P^2 \right )\over \big(V[X^{-1}_2] \big)^2 r^{2}}\Bigg]\m{d}t^2\;
\nonumber \\
& & + \;{\m{d}r^2\over {\dfrac{r^2}{L^2}}\;-\;1\;-\;{\dfrac{8\pi M}{ V[X^{-1}_2] r}}\;+\;{\dfrac{4\pi \left (-\,Q^2 + P^2 \right )}{ \big(V[X^{-1}_2] \big)^2 r^{2}}}} \;+\; r^2\,\m{d}\Omega^2[X_2^{-1}],
\end{eqnarray}
where the ``dy'' in $g^E(\m{AdSdyRN^{-1}_{4})}$ henceforth denotes ``dyonic''.

The quantity $\mathfrak{S^{\m{E}}}(r)$ in (\ref{B}) can be computed directly for this geometry, taking the surfaces $\Sigma$ to be the surfaces $r = \text{constant}$ (in which the coordinates are those of the Riemann surface, together with Euclidean ``time'' $t$). Up to an overall positive multiplicative factor\footnote{As usual, Euclidean ``time'' will always be circular throughout this work; the multiplicative factor here includes the circumference of this circle. This comment applies to all of our examples.}, it turns out to be given by
\begin{equation}\label{D}
\mathfrak{S^{\m{E}}}(\m{AdSdyRN^{-1}_{4}})(r)\;=\;K\;-{r\;+\;{\dfrac{8\pi M}{V[X^{-1}_2]}}\;-\;{\dfrac{4\pi \left (-\,Q^2 + P^2 \right )}{ \big(V[X^{-1}_2]\big)^2r}}\over \sqrt{{\dfrac{1}{L^2}}\;-\;{\dfrac{1}{r^2}}\;-\;{\dfrac{8\pi M}{V[X^{-1}_2] r^3}}\;+\;{\dfrac{4\pi \left (-\,Q^2 + P^2 \right )}{\big(V[X^{-1}_2]\big)^2r^4}}}+{\dfrac{1}{L}}},
\end{equation}
where $K$ is a positive constant (which depends on the black hole parameters). This is positive for \emph{some} surfaces $\Sigma$ but not for all: certainly not for those which are far from the black hole. The conditions (\ref{A}) and (\ref{B}) are violated, so holography simply does not work here.

However, this is not alarming: while this geometry is perfectly well-behaved classically, it is not well-behaved when embedded in string theory. For the Lorentzian version of the quantity defined by equation (\ref{B}), let us call it $\mathfrak{S^{\m{L}}}(\m{AdSdyRN^{-1}_{4}})(r)$, is, up to factors involving the brane tension, precisely the action of a BPS brane in the Lorentzian black hole geometry \cite{kn:seiberg}. This function behaves in much the same way as $\mathfrak{S^{\m{E}}}(\m{AdSdyRN^{-1}_{4}})(r)$, that is\footnote{$\mathfrak{S^{\m{L}}}(\m{AdSdyRN^{-1}_{4}})(r)$ is obtained by reversing the coefficient of $Q^2$ in (\ref{D}) and adjusting the value of $K$ accordingly (one can show that it remains positive). Both functions behave in much the same way, because the dominant term in the numerator on the right side of (\ref{D}) is the one involving $r$, and the signs of the other terms are irrelevant when $r$ is large.}, at sufficiently large $r$ it becomes smaller than its value at the event horizon. The black hole will therefore generate arbitrary quantities of branes by a Schwinger-like process \cite{kn:maldacena, kn:KPR} at such values of $r$, and these branes will have no tendency to contract back into the hole. The implicit assumption that the spacetime is static is therefore not tenable. \emph{The absence of this ``Seiberg-Witten instability'' is the Lorentzian interpretation of ``consistency''} in the sense of \cite{kn:ferrari3}.

Holography implies that the boundary field theory must be pathological in this case. In fact, however, we do not need holography to see this: Seiberg and Witten explain that the field theory on the boundary here is ill-defined, because the scalar curvature at infinity is evidently negative, and this causes a certain scalar to acquire an effective negative squared mass. In short, the failure of condition (\ref{A}) in this example is not a matter for concern; in any case, the geometry at infinity in this case is not of any great interest for applications.

This discussion prompts the question, however: is this kind of example, with negative scalar curvature at infinity (so that holography is in any case superfluous), the \emph{only} kind of example in which holography is inconsistent? In view of the Lorentzian interpretation we have just discussed, that appears to be a reasonable conjecture, and an attempt to confirm it was the subject of \cite{kn:ferrari3}. In that work, certain results from the mathematical literature \cite{kn:lee,kn:wang,kn:wang2} (see also \cite{kn:galloway}) were extended to show that the inequality (\ref{B}) does hold when the scalar curvature at infinity is not negative (more precisely, if the Yamabe invariant of the underlying manifold is non-negative) provided that one confines attention to \emph{Euclidean} bulk manifolds which are \emph{Einstein} manifolds like AdS itself. In the simplest cases, then, holography is consistent if one can justify working in Euclidean signature and avoids negatively curved manifolds at infinity: in particular, all is well in what is by far the most important case, when the boundary is (globally) conformally flat.

Unfortunately, in applications of holography one is not exclusively or even primarily interested in the ``simplest cases'', that is, in an Einstein bulk. For example, in the application to the quark-gluon plasma \cite{kn:solana,kn:pedraza,kn:youngman,kn:gubser,kn:janik}, one needs an electric field in the bulk (and the corresponding back-reacted black hole metric) to deal with a non-zero baryonic chemical potential in the field theory, and in other applications (for example, \cite{kn:hartkov,kn:dyon,kn:82,kn:83}) one similarly needs a fully back-reacted magnetic field in the bulk; in neither case is the bulk an Einstein manifold. One has therefore to ask whether the mathematical results used in \cite{kn:ferrari3} can be generalised to deal with this situation. As we shall discuss, such theorems do exist, but they do not help us here.

More surprisingly, perhaps, the assumption that the bulk geometry is Euclidean can also lead to difficulties. As we have seen, the positivity of the quantity $\mathfrak{S^{\m{E}}}$ in (\ref{B}) has a very explicit Lorentzian interpretation in terms of the demand that a certain brane action should not drop below its value at the event horizon of a black hole in the bulk. But as it is defined by the geometry of the bulk, $\mathfrak{S}$ depends on the black hole parameters, and these may differ between the Euclidean and Lorentzian versions of the geometry. This difference can affect the asymptotic behaviour of $\mathfrak{S}$; and it can happen that the Euclidean version is well-behaved \emph{while the Lorentzian version is not}\footnote{We are not aware of any examples in which the reverse is the case, that is, in which the Lorentzian behaviour is better than the Euclidean; one suspects that this is not possible. Of course, it is possible for both to behave badly, as we saw above.}. It will turn out that this problem can arise even when the bulk \emph{is} an Einstein manifold.

Now, as we have seen, ``holographic inconsistency'' has a different interpretation in the Lorentzian and Euclidean cases. In the latter, it is a genuine mathematical inconsistency and is therefore completely unacceptable. However, the Lorentzian version is associated with an instability, which takes time to develop ---$\,$ and, in applications, one is sometimes dealing with an extremely short-lived system (like the quark-gluon plasma formed by a heavy-ion collision), which may not survive for a time sufficient for the instability to be relevant. We can therefore state the case as follows: we consider that holography is only fully consistent when
\begin{itemize}
\item[(a)] the Euclidean version is well-behaved, \emph{and}
\item[(b)] the Lorentzian version is also well-behaved, unless the system exists for a sufficiently short period that any Lorentzian misbehaviour would not have sufficient time to disturb the entire bulk geometry, particularly the vicinity of the event horizon in the case of a black hole bulk spacetime.
\end{itemize}

Granting all this, we see that considerations like those of \cite{kn:ferrari3}, assuring us that the Euclidean quantity will always behave well when the bulk is Einstein and the Yamabe invariant at infinity is non-negative, are important and necessary, but not sufficient to settle the question of consistency.

We will see that the question we have raised ---$\,$ is holography ever inconsistent in a ``physically reasonable'' situation? ---$\,$ has no straightforward answer; that is, there is no simple algorithm for deciding the question. We have to resort to a case-by-case consideration, and, in this work, we shall consider several concrete examples (all involving applications of holography to the quark-gluon plasma). The remarkable conclusion we reach is that, despite the apparent absence of any general way of determining whether holography is consistent in the sense of \cite{kn:ferrari3}, it \emph{is} consistent in all of these very diverse examples. This seeming coincidence suggests that the consistency of holography, expressed in this context by equation (\ref{A}), has the character of a ``law of physics'' (as opposed to a mere technical question that can be settled purely mathematically).

Because the mathematics is more straightforward, it will be simpler to begin with the non-Einstein case.
\addtocounter{section}{1}
\section* {\large{\textsf{2. The Non-Einstein Case I: Dyonic Planar AdS Black Holes}}}
We begin with a ``dyonic'' four-dimensional black hole with a \emph{planar} \cite{kn:lemmo,9607071,9609065,9808032,kn:dyon} event horizon, the planar case being the one of greatest interest for applications. The four-dimensional asymptotically AdS dyonic Reissner-Nordstr\"om metric with planar (zero-curvature) event horizon takes the form
\begin{eqnarray}\label{E}
g(\m{AdSdyRN^{0}_{4})} & = & -\, \Bigg[{r^2\over L^2}\;-\;{8\pi M^*\over r}+{4\pi (Q^{*2}+P^{*2})\over r^2}\Bigg]\m{d}t^2\; \nonumber \\
& &  + \;{\m{d}r^2\over {\dfrac{r^2}{L^2}}\;-\;{\dfrac{8\pi M^*}{r}}+{\dfrac{4\pi (Q^{*2}+P^{*2})}{r^2}}} \;+\;r^2\Big[\m{d}\psi^2\;+\;\m{d}\zeta^2\Big],
\end{eqnarray}
where $\psi$ and $\zeta$ are dimensionless planar coordinates, $L$ is the asymptotic AdS curvature radius, and $M^*$, $Q^*$, and $P^*$ are geometric parameters related to the mass and electric and magnetic charges per unit horizon area. (Such parameters are necessary here in case the event horizon is truly planar, that is, not compactified.) For all values of the parameters, the spacetime at infinity can be interpreted as a three-dimensional spacetime (signature $(-\,+\,+)$) embedded in a flat spacetime (in the application to heavy-ion collisions) or in a globally conformally flat spacetime (such as an FRW cosmology with flat spatial sections; the boundary then corresponds to the three-dimensional spacetime associated with a specific two-dimensional plane, which, by isotropy and homogeneity, can be chosen arbitrarily).

The event horizon is located at $r = r_h$, given by a solution of
\begin{equation}\label{F}
{r_h^2\over L^2}\;-\;{8\pi M^*\over r_h}+{4\pi (Q^{*2}+P^{*2})\over r_h^2} = 0.
\end{equation}
This quartic has a positive real solution (that is, cosmic censorship is ensured) provided that
\begin{equation}\label{G}
\left (P^{*2}+Q^{*2}\right )^3 \leqslant \frac{27}{4}\pi M^{*4}L^2.
\end{equation}

The metric $g(\m{AdSdyRN^{0}_{4})}$ is not an Einstein metric if either charge is non-zero; the Ricci tensor has two terms, one proportional to the metric, the other to the stress-energy-momentum tensor associated (outside the black hole) with a (dimensionless) electromagnetic potential form given by\footnote{Strictly speaking, one needs to fix the gauge in such expressions (by adding certain constants to the components of $A$) so that the Euclidean version of the potential form is regular everywhere. We shall dispense with these constant terms here and henceforth, since they play no role in this work.}
\begin{equation}\label{H}
A\;=\;-\,{Q^*\over r L}\m{d}t\;+\;{P^*\psi\over L} \m{d}\zeta.
\end{equation}

Equation (\ref{H}) is of basic importance here, because it teaches us how to perform the continuation to the Euclidean domain. Since $t$ is complexified ($t\,\rightarrow \,it$) but $\psi$ and $\zeta$ are not, it is clear that $Q^*$ must be complexified ($Q^*\,\rightarrow \,-\,iQ^*$) \emph{but $P^*$ must not}. The Euclidean version of $g(\m{AdSdyRN^{0}_{4})}$, let us call it $g^E(\m{AdSdyRN^{0}_{4})}$, therefore takes the form
\begin{eqnarray}\label{I}
g^E(\m{AdSdyRN^{0}_{4})} & = &  \Bigg[{r^2\over L^2}\;-\;{8\pi M^*\over r}+{4\pi (-\,Q^{*2}+P^{*2})\over r^2}\Bigg]\m{d}t^2\; \nonumber \\
& &  + \;{\m{d}r^2\over {\dfrac{r^2}{L^2}}\;-\;{\dfrac{8\pi M^*}{r}}+{\dfrac{4\pi (-\,Q^{*2}+P^{*2})}{r^2}}} \;+\;r^2\Big[\m{d}\psi^2\;+\;\m{d}\zeta^2\Big].
\end{eqnarray}
A ``Euclidean event horizon'' (which of course is just a regular point in the $t-r$ plane, a particular copy of the two-dimensional space parametrised by $\psi$ and $\zeta$), located at $r = r_h^E$, can be defined in the obvious way; by reversing the coefficient of $Q^{*2}$, one sees from the inequality ({\ref{G}) that, if the Lorentzian event horizon exists, then so must the Euclidean version.

In equation (\ref{I}), one should think of $t$ as being compactified, and it will be convenient also to compactify $\psi$ and $\zeta$. The topology at infinity is then that of a three-dimensional torus, which has zero Yamabe invariant; by the theorem of Wang \cite{kn:wang, kn:wang2}, therefore, if this manifold were an Einstein manifold, the quantity $\mathfrak{S}$ would have to be non-negative and then, by the argument in \cite{kn:ferrari3}, condition (\ref{A}) would have to be satisfied.

Since the manifold is not Einstein, we need to consult the mathematical literature to determine whether Wang's result can be generalised accordingly. Such results were discussed by Witten and Yau \cite{kn:wittenyau} and improved by Cai and Galloway \cite{kn:galloway}, and are discussed from a physics point of view in \cite{kn:trieste}. The details are complicated, but can be summarized as follows.

If a (four-dimensional) Riemannian asymptotically hyperbolic\footnote{The ``asymptotically hyperbolic'' condition is the Euclidean analogue of ``asymptotically AdS''.} manifold is Einstein, then the eigenvalues of the Ricci tensor are all equal to $-3/L^2$. If it is not Einstein, then the eigenvalues are functions of position which merely \emph{approach} $-3/L^2$ towards infinity. If we denote the eigenvalue functions by $R_j$, then results generalizing Wang's theorem \cite{kn:wang, kn:wang2} can be obtained provided that, for each $j$, $R_j\,+\,(3/L^2) \;\geqslant \;0$ everywhere and provided that the functions $R_j\,+\,(3/L^2)$ approach zero ``sufficiently quickly'' towards the boundary; the reader can obtain a more precise statement from \cite{kn:galloway}.

Unfortunately, a detailed calculation given in \cite{kn:trieste} shows that, in the case of an electromagnetic field in the bulk (or any other gauge field), while all of the functions $R_j\,+\,(3/L^2)$ do approach zero sufficiently quickly towards infinity, not all of them can be non-negative in the Euclidean case. Thus, these generalised theorems do not help us; and in fact we will see that there can be no straightforward mathematical statement in this case.

Instead, let us resort to a direct calculation of $\mathfrak{S^{\m{E}}}$ for this metric. This was done in the Lorentzian case in \cite{kn:83}; the Euclidean version is then, again up to an overall positive factor,
\begin{equation}\label{J}
\mathfrak{S^{\m{E}}}(\m{AdSdyRN^{0}_{4}})(r)\;=\; { \left (-8\pi M^* + {\dfrac{4\pi (-\,Q^{*2} + P^{*2})}{r}} \right )/L\over 1+\sqrt{1-{\dfrac{8\pi M^*L^2}{r^3}}+{\dfrac{4\pi (-\,Q^{*2} + P^{*2})L^2}{r^4}}}}+{(r^E_h)^3\over L^3}.
\end{equation}
Notice that $\mathfrak{S^{\m{E}}}(\m{AdSdyRN^{0}_{4})}(r)$ vanishes at the Euclidean event horizon, that being the origin of coordinates in the $t-r$ plane. (The related Lorentzian brane action must likewise vanish at the Lorentzian horizon.) It is easy to see that it is then positive as $r$ increases, but its subsequent behaviour is less easy to guess.

In fact, a straightforward analysis (see \cite{kn:83} for the details in the Lorentzian case) shows that this function is always positive if and only if
\begin{equation}\label{K}
4\pi (P^{*2}\,-\,Q^{*2})L^2 \;\leqslant \;(r^E_h)^4.
\end{equation}
It is clear that this will automatically hold in some cases, but we will see that it sometimes does not: thus indeed it is the case that, when the bulk is not an Einstein manifold, condition (\ref{B}) can be violated even though the Yamabe invariant at infinity is not negative here. In those cases, we need a physical interpretation of this fact, so that we can judge whether holography fails to be consistent in a physically realisable situation.

At this point it is convenient to focus on the two simplest special cases of this inequality, treating them separately.

\subsubsection*{{\textsf{2.1 The Purely Magnetic Case }}}
We begin with a purely magnetically charged black hole, $Q^* = 0$; the holographic dual is a plasma with zero or negligible baryonic chemical potential (see below), permeated by a transverse magnetic field with intensity related to $P^*$ (see equation (\ref{H}) above). In this case, the Euclidean and Lorentzian versions of the quantity $\mathfrak{S}$ coincide, so we can state the physical meaning of the inequality (\ref{K}) by quoting directly from the conclusions of \cite{kn:82}, where this case was investigated in detail: one finds in this case that (\ref{K}) is equivalent to
\begin{equation}\label{L}
B\;\leqslant \;2\pi^{3/2}T^2,
\end{equation}
where $B$ is the magnetic field strength (defined in terms of the flux through a two-dimensional surface) associated with the boundary theory, and $T$ is the Hawking temperature of the black hole (and therefore, by holography, the temperature of the boundary field theory). Thus, if we accept the claim that the boundary field theory resembles a strongly coupled quark-gluon plasma with negligible baryonic chemical potential but subjected to a transverse magnetic field, the fundamental consistency condition (\ref{A}) forbids the magnetic field to be extremely large relative to the squared temperature.

It is interesting that an abstract consistency condition can be related in such a direct manner to the physical parameters of a quasi-realistic physical system; in fact, the consistency condition is making an unexpected physical prediction, that the inequality (\ref{L}) must hold if the plasma does admit a holographic description. Notice in this connection that this case is quite unlike the one, considered earlier, with a bulk containing a black hole with a negatively curved event horizon; for, in that case, we did not need holography to inform us that the boundary theory might be pathological: that was clear from the coupling of a certain scalar to the boundary scalar curvature. Here the boundary has zero scalar curvature, indeed it is flat, so holography reveals something genuinely novel.

But do real plasmas actually satisfy (\ref{L})? As a matter of fact, it is well known that gigantic magnetic fields (of the order $10^{17}$ gauss or more) do arise in two actual quark plasma systems: in the plasma generated by peripheral collisions at heavy-ion colliders \cite{kn:skokov,kn:kharzeev1,kn:naylor,kn:kharzeev2}, and during the plasma era of the early Universe \cite{kn:reviewA,kn:reviewB,kn:planck}. In both cases, however, the temperature is also enormous, so the status of (\ref{L}) is unclear.

If we consider a plasma temperature around the hadronization temperature (say, 150 MeV) which is a natural choice\footnote{In the case of cosmic magnetic fields, the choice of temperature is not important because $B$ and $T^2$ are normally assumed to evolve in the same way with the cosmic expansion, so (\ref{L}) will be satisfied at all temperatures if it is satisfied at any given temperature during the plasma era.} in view of the possibility that the plasma may be less strongly coupled at significantly higher temperatures, then (\ref{L}) takes the explicit form
\begin{equation}\label{M}
eB\;  \lesssim  \; 3.6 \times 10^{18}\;\; \text{gauss}.
\end{equation}
Interestingly, this is just above the estimated maximal magnetic fields attained in collisions at the RHIC facility \cite{kn:skokov}. The ALICE facility at the LHC \cite{kn:ALICE} observes the plasma at about twice this temperature, but also at significantly higher values of $B$, and it is possible that the ALICE plasma comes very close to saturating (\ref{L}).

However, it may be premature to attach much importance to this, for several reasons. Most importantly, the magnetic fields generated in heavy ion collisions are very short-lived, perhaps too short-lived for the Seiberg-Witten instability (discussed in the preceding section) to affect them \cite{kn:77}; also, they are associated with extremely large angular momentum densities, which we are not taking into account here, though it is known that they are important holographically \cite{kn:klemm,kn:shear,kn:79}. Let us turn, then, to the case of the cosmic plasma, where these complications do not normally (but see \cite{kn:brand}) arise\footnote{The cosmic plasma endures for a period of time (several microseconds) which is very long by strong-interaction standards, so, if it violates (\ref{L}), there will be more than sufficient time for the corresponding instability to develop; in other words, the system would then be inconsistent in both the Euclidean and Lorentzian senses.}.

In this case, a huge magnetic field might be present at the end of the cosmic plasma era, surviving from certain effects during the Inflationary era: see \cite{kn:reviewA,kn:reviewB}. However, the maximal estimated size of the magnetic field (at hadronization) in this case is \cite{kn:82,kn:83} around $3.7 \times 10^{17}$ gauss, which still satisfies (\ref{L}).

To summarize: holographic consistency (in both senses) can, in principle, be violated by a concrete physical system with a non-Einstein holographic dual: a relatively long-lived quark-gluon plasma inhabiting a flat (or conformally flat) spacetime, accompanied by a sufficiently large magnetic field. In fact, however, \emph{no known example of such a plasma has a magnetic field which clearly violates the inequality (\ref{L}).}

\subsubsection*{{\textsf{2.2 The Purely Electric Case }}}
In contrast to the magnetic case, in the purely electric case ($P^* = 0$), inequality (\ref{K}) is clearly automatically satisfied for all values of $Q^*$. The case $P^* = 0$ in $g^E(\m{AdSdyRN^{0}_{4})})$ therefore provides us with a concrete example of a non-Einstein, Euclidean bulk in which the condition (\ref{B}) \emph{is} indeed satisfied everywhere. That is, a non-Einstein bulk does not necessarily violate (\ref{B}).

If we consider only the Euclidean case, we will conclude that holographic consistency always holds in this case. But if we require also that the Lorentzian version of the system should be well-behaved, then we have to require (from (\ref{K})) that
\begin{equation}\label{N}
4\pi Q^{*2}L^2 \;\leqslant \;(r_h)^4,
\end{equation}
where $r_h$ is now the coordinate of the Lorentzian event horizon, and it is no longer clear that this will always hold.

The implications of imposing (\ref{N}) in the Lorentzian case were explored in \cite{kn:83}. As is well known \cite{kn:koba}, the electric charge of the black hole is related holographically to the baryonic chemical potential of the boundary field theory, $\mu_B$. The relation is however not the same as the relation of $P^*$ to the boundary magnetic field, so the physical interpretation takes a quite different form. As in the previous section, and for the same reasons, it is difficult to apply our results to the case of the plasma produced in heavy-ion collisions; in any case, such plasmas normally have very low values of $\mu_B$, certainly at ALICE \cite{kn:ALICE}. (The next phase of the beam scan experiments at RHIC \cite{kn:STAR,kn:BEAM}, and future experiments at FAIR \cite{kn:FAIR}, are expected to change this situation, however.)

We therefore turn again to the cosmic plasma, assuming as usual that it resembles the boundary field theory. Here too, the conventional description involves a plasma with a very low value of $\mu_B$; but recently a new theory of the evolution of the cosmic plasma has been suggested, the ``Little Inflation'' theory \cite{kn:tillmann1,kn:tillmann2,kn:tillmann3}. In this approach, $\mu_B/T$, where $T$ is the temperature of the plasma, can be quite large, well above unity. This is reconciled with the observed baryon asymmetry by postulating that the end of the plasma era is triggered by the decay of a false QCD vacuum, associated with a first-order phase transition to the hadronic state. This is an interesting new approach to early universe cosmology, and it is possible that some of its many concrete predictions may be confirmed in the reasonably near future.

One expects the inequality (\ref{N}) to be relevant to this theory, and so indeed it is: in \cite{kn:83} it is shown that (\ref{N}) is equivalent to the restriction
\begin{equation}\label{O}
\mu_B/T \; \leqslant \; \left(1\;-\;2^{1/3}\;+\;2^{2/3} \right)\sqrt{\pi}\;\approx 2.353.
\end{equation}
This is a very strong condition in the ``Little Inflation'' context, and, as explained in \cite{kn:83}, it forces the cosmic plasma to hadronize quite close to the quark matter critical point \cite{kn:race}, where very distinctive phenomena analogous to critical opalescence \cite{kn:csorgo} may soon be observed in the beam scan experiments mentioned earlier. It is not hard to imagine that such fluctuation phenomena might prove to be irreconcilable with cosmological observations; in which case one \emph{might} eventually be led to conclude that ``Little Inflation'' actually involves values of $\mu_B/T$ well above 2.353. (This would \emph{not} be a problem for ``Little Inflation'' itself, where values of $\mu_B/T$ far larger than this ---$\,$ up to $\approx 100$ ---$\,$ are quite acceptable.) In short, it is perfectly conceivable that near-future observations will indicate that \emph{Lorentzian} holographic consistency is violated in ``Little Inflation'' cosmology, although \emph{Euclidean} holographic consistency always holds.

However, at present there are many unknowns here: for example, even the location of the critical point in the quark matter phase diagram remains controversial \cite{kn:exploring}, and of course ``Little Inflation'' has itself yet to be confirmed. At present, then, \emph{ we have no convincing evidence to suggest that (\ref{O}) is violated}, even though, in principle, it might be: see \cite{kn:83} for the details.

To summarize in a manner parallel to the summary at the end of the preceding section: \emph{the Lorentzian version} of the consistency condition (\ref{A}) can, in principle, be violated by a concrete physical system with a non-Einstein holographic dual: a relatively long-lived quark-gluon plasma inhabiting a flat (or conformally flat) spacetime, described by a sufficiently large baryonic chemical potential. In fact, however, \emph{no known example of such a plasma has a baryonic chemical potential which clearly violates the inequality (\ref{O}).}

Summarizing this entire Section: in the case of a non-Einstein bulk, one has no guarantee that Euclidean holographic consistency will be satisfied, even if the conformal boundary has zero Yamabe invariant; and one has a concrete example (the purely magnetic case, above) where it is not satisfied, \emph{but} this requires magnetic fields stronger than any confirmed actually to exist. On the other hand, one also has a concrete example (the purely electric case) in which Euclidean holographic consistency does hold for all values of all parameters, yet in which Lorentzian consistency can fail in principle: but, once again, no known system actually does cause it to fail.

\subsection*{{\textsf{3. The Non-Einstein Case II: Scalars in the Bulk}}}
Electromagnetic fields are of course but one way of causing the bulk to be non-Einstein. Another form of bulk matter important in holographic applications is defined by \emph{scalar} (dilaton) fields. These are important in, for example, the holographic theory of the quark matter equation of state: see \cite{kn:dilaton} for a recent example with many references.

As in the example at the end of the preceding section, one is interested here in electrically charged\footnote{For the sake of simplicity, in this section we consider only electrically charged black holes. The magnetic case can be studied using electromagnetic duality: bear in mind, however, that the \emph{string} metric transforms non-trivially, because under the electromagnetic duality the dilaton transforms according to $\phi \rightarrow -\phi$. The dyonic solution is non-trivial even in the asymptotically flat case, since the presence of both electric and magnetic charges necessitates the presence of an (antisymmetric 3-form) axion field. See \cite{kn:horowitz, kn:GHS}.} dilatonic black holes in an AdS background. A generic scalar potential leads to black holes which are \emph{not} asymptotically AdS in the strict sense \cite{9407021,9412076,9202031}; we will confine ourselves here to black holes which \emph{are} asymptotically AdS. These are the Gao-Zhang black holes \cite{kn:gz}. These too have important applications; for example they have recently been used to good effect in the holographic theory of the thermalization of the quark plasma \cite{kn:zhang}; this is in fact one of the most active areas of applied holography.

The construction of the Gao-Zhang black holes is highly nontrivial: Gao and Zhang were forced to use a combination of \emph{three} Liouville-type potentials.
The corresponding action in $n$-dimensional spacetime is
\begin{equation}
S=-\frac{1}{16\pi}\int \m{d}^nx \sqrt{-g} \left[R -\frac{4}{n-2}(\nabla \phi)^2 - V(\phi) - e^{-\frac{4\alpha \phi}{n-2}}F^2\right], ~\alpha \geqslant 0,
\end{equation}
where $\alpha$ is the coupling of the dilaton to the electromagnetic field.

The exact form of the potential is rather complicated, and is not important for our discussion here. In the case $\alpha=0$, the potential reduces to the (negative) cosmological constant,
and the dilaton field is identically zero (see for example equation (5) of \cite{kn:zhang}).
Note that, in many applications of holography, especially if one is only interested in the IR physics, then the precise details of the potential are not required for determining the low-energy behavior arising from the near-horizon geometry. This allows one to work with an effective action with its corresponding approximate black hole solution. However, for the purpose for analyzing Seiberg-Witten instability, such an effective action is not suitable since branes are sensitive to the global geometry of the spacetime. In other words, we wish to study the brane action not only for the near-horizon region, but at \emph{all} values of coordinate radius $r$. Therefore we confine our attention to an exact solution of Gao-Zhang type.

The Gao-Zhang black hole solution \cite{kn:gz} (or, in our terminology, the $n$-dimensional asymptotically AdS dilatonic Reissner-Nordstr\"om black hole) is of the form
\begin{equation}
g(\m{AdSdilRN}^{k}_{n})=-U(r)\m{d}t^2 + W(r) \m{d}r^2 + [f(r)]^2 \m{d}\Omega^2[X^k_{n-2}],
\end{equation}
where $d\Omega^2$ is a (dimensionless) metric on $X^k_{n-2}$, a $(n-2)$-dimensional Riemannian manifold of constant curvature $k$.
The coefficient functions are
\begin{equation}
\begin{cases}
U(r) = \left[k-\left(\dfrac{c}{r}\right)^{n-3}\right]\left[1-\left(\dfrac{b}{r}\right)^{n-3}\right]^{1-\gamma(n-3)} + \dfrac{r^2}{L^2}\left[1-\left(\dfrac{b}{r}\right)^{n-3}\right]^\gamma,
\\
\\
W(r) = U(r)^{-1}\left[1-\left(\dfrac{b}{r}\right)^{n-3}\right]^{-\gamma (n-4)}, \\
\end{cases}
\end{equation}
and
\begin{equation}
f(r)^2 = r^2\left[1-\left(\frac{b}{r}\right)^{n-3}\right]^\gamma, ~~~\gamma = \frac{2\alpha^2}{(n-3)(n-3+\alpha^2)},
\end{equation}
where $\gamma$ is of course unrelated to the constant in equation (\ref{A}), and where $b$ and $c$ are constants related to the physical mass and electric charge by the equations (see \cite{1002.0202})
\begin{equation}
M=\frac{V[X^k_{n-2}]}{16\pi}(n-2)\left[c^{n-3} + kb^{n-3}\left(\frac{n-3-\alpha^2}{n-3+\alpha^2}\right)\right], ~~\text{and}
\end{equation}
\begin{equation}
Q=\frac{V[X^k_{n-2}]}{4\pi}\left[\frac{(n-2)(n-3)^2}{2(n-3+\alpha^2)}(bc)^{n-3}\right]^{\frac{1}{2}},
\end{equation}
where $V[X^k_{n-2}]$ is the dimensionless volume of $X^k_{n-2}$.

In four dimensions, and for zero spatial curvature $k=0$, we have
\begin{equation}
U(r)=-\frac{c}{r}\left[1-\frac{b}{r}\right]^{\frac{1-\alpha^2}{1+\alpha^2}} + \frac{r^2}{L^2} \left[1-\frac{b}{r}\right]^{\frac{2\alpha^2}{1+\alpha^2}},
\end{equation}
and $W(r)=U(r)^{-1}$. Note that for this class of AdS black holes, $g_{tt}g_{rr}=-1$ only holds in four dimensions\footnote{It is possible to use a coordinate system $(t, R, \psi, \zeta)$, in which $R$, \emph{unlike} $r$, is an areal radius, and $\psi, \zeta$ are coordinates on a flat space. But then  $g_{tt}g_{RR}\neq -1$ even in four dimensions. See \cite{ted}.}.
We also have
\begin{equation}
f(r)^2 = r^2\left(1-\frac{b}{r}\right)^{\frac{2\alpha^2}{1+\alpha^2}}.
\end{equation}

The mass and charge density parameters are given by
\begin{equation}
M^* = \frac{M}{V[X^0_{2}]} = \frac{c}{8\pi}, ~~ Q^*=\frac{Q}{V[X^0_{2}]}=\frac{1}{4\pi}\left(\frac{bc}{1+\alpha^2}\right)^{\frac{1}{2}}.
\end{equation}
Thus, under Wick rotation to Euclidean signature, $Q^{*2} \rightarrow -Q^{*2}$ implies $b \rightarrow -b$.

The (Euclidean) quantity $\mathfrak{S^{\m{E}}}$ for this metric, up to the usual positive constant factor, is
\begin{flalign}
\mathfrak{S^{\m{E}}}(\m{AdSdilRN^{0}_{4}})(r) &= r^2\left[1+\frac{b}{r}\right]^{\frac{2\alpha^2}{1+\alpha^2}} \left[\frac{r^2}{L^2}\left(1+\frac{b}{r}\right)^{\frac{2\alpha^2}{1+\alpha^2}} -\frac{c}{r}\left(1+\frac{b}{r}\right)^{\frac{1-\alpha^2}{1+\alpha^2}}\right]^{\frac{1}{2}}\\ \notag &~~~~-\frac{3}{L}\int_{r_h^E}^r s^2\left[1+\frac{b}{s}\right]^{\frac{2\alpha^2}{1+\alpha^2}} \m{d}s \notag \\
&= \frac{r^3}{L}\left[1+\frac{b}{r}\right]^{\frac{3\alpha^2}{1+\alpha^2}}\left[1-\frac{cL^2}{r^3}\left(1+\frac{b}{r}\right)^{\frac{1-3\alpha^2}{1+\alpha^2}}\right]^{\frac{1}{2}} \\ \notag &~~~~-\frac{3}{L}\int_{r_h^E}^r s^2\left[1+\frac{b}{s}\right]^{\frac{2\alpha^2}{1+\alpha^2}} \m{d}s,
\end{flalign}
where $r_h^E$ is the value of $r$ at the ``Euclidean horizon''.

Since the action is rather complicated, and since in general there is no closed form expression for $r_h^E$, let us fix $r_h^E=1$ in some unit of length. This fixes the relation between the parameters $b$ and $c$, and because the Euclidean horizon satisfies
\begin{equation}
1-\frac{cL^2}{(r_h^E)^2}\left(1+\frac{b}{r_h^E}\right)^{\frac{1-3\alpha^2}{1+\alpha^2}}=0,
\end{equation}
we have $(1+b)^\frac{3\alpha^2-1}{1+\alpha^2}=cL^2$. The action is now
\begin{flalign}
\mathfrak{S^{\m{E}}}(\m{AdSdilRN^{0}_{4}})(r) =&\frac{r^3}{L}\left[1+\frac{b}{r}\right]^{\frac{3\alpha^2}{1+\alpha^2}}\left[1-\frac{(1+b)^{\frac{3\alpha^2-1}{\alpha^2+1}}}{r^3}\left(1+\frac{b}{r}\right)^{\frac{1-3\alpha^2}{1+\alpha^2}}\right]^{\frac{1}{2}} \\ \notag &-\frac{3}{L}\int_{1}^r s^2\left[1+\frac{b}{s}\right]^{\frac{2\alpha^2}{1+\alpha^2}} \m{d}s.
\end{flalign}

A numerical investigation indicates that the Euclidean action is always positive (an example is provided in Figure (\ref{fig1})). In fact, for large $r$, by expanding in powers of $r$, it can be shown that $\mathfrak{S^{\m{E}}}(\m{AdSdilRN^{0}_{4}})(r)$ grows linearly in $r$. Specifically, we have asymptotically,
\begin{equation}\label{OYC}
\mathfrak{S^{\m{E}}}(\m{AdSdilRN^{0}_{4}})(r) \sim \frac{3b^2\alpha^2}{2L(1+\alpha^2)^2} r + \text{const}(\alpha).
\end{equation}
Here, the term $\text{const}(\alpha)$ is an $\alpha$-dependent constant. For $\alpha=0$ (the electrically charged AdS-Reissner-Nordstr\"om case), the constant term is the only term that survives as $r$ tends to infinity, and it is positive.

Recall that for AdS Reissner-Nordstr\"om black holes with flat event horizons but no scalars or magnetic charges, this quantity is likewise everywhere positive. Thus, the introduction of the dilaton does not change the situation, for any value of the dilaton coupling.

\begin{figure}[!h]
\centering
\includegraphics[width=6.0in]{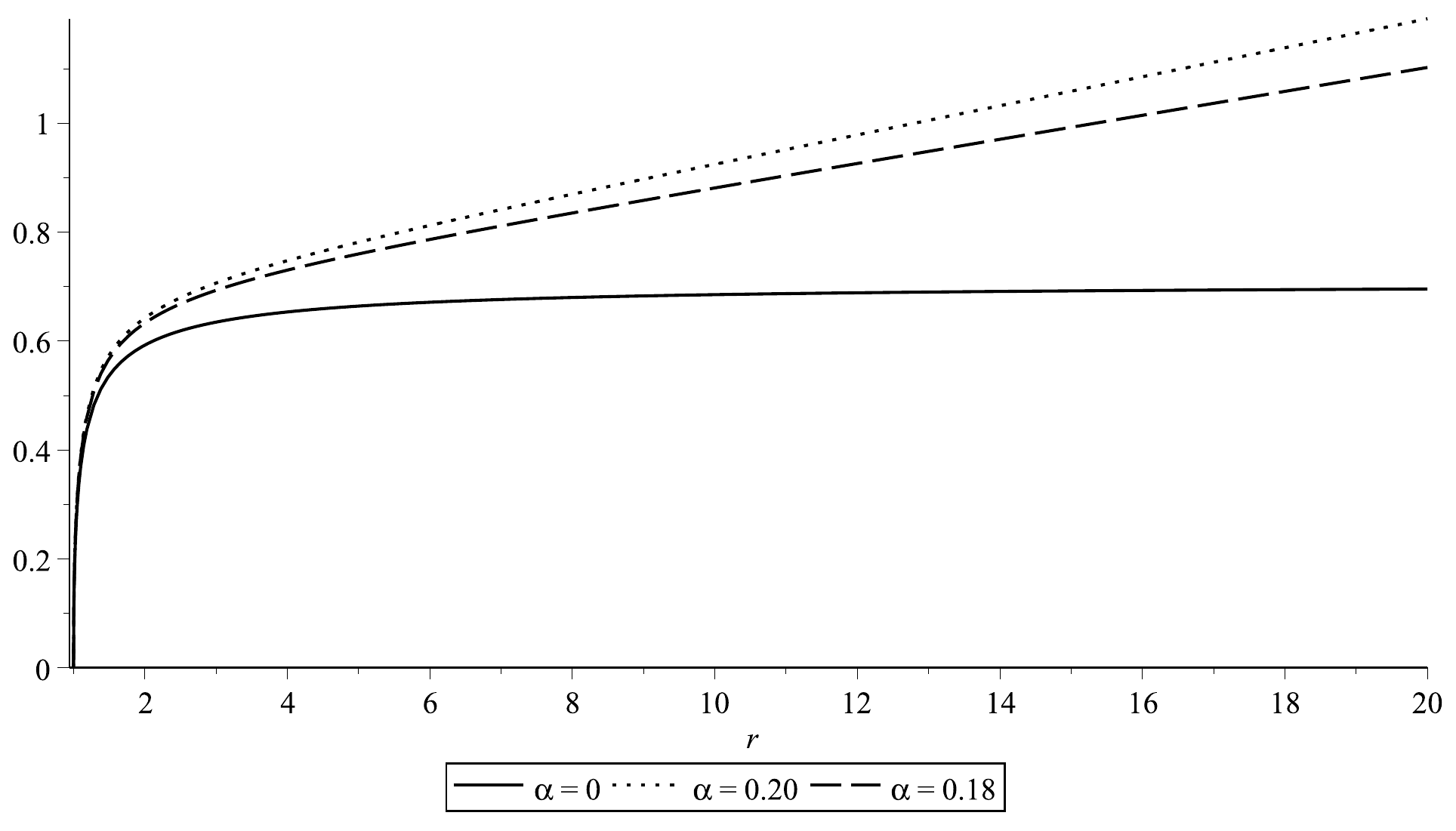}
\caption{\label{fig1}
The Euclidean brane actions $\mathfrak{S}^\text{E}(r)$ of the Gao-Zhang black hole (up to a positive factor) with various values of the dilaton coupling parameter $\alpha$, the same charge parameter $b=0.7$, and horizons held fixed at $r_h^E=1$. Here we set $L=1$, however the value of $L$ only contributes an overall factor to the action, and thus does not affect its positivity. }
\end{figure}

It is otherwise in the Lorentzian case, however (although in both Euclidean and Lorentzian cases, the effect of the dilaton field is to increase the value of the brane action). This case was investigated at length (in five dimensions, but the four-dimensional situation is similar) in \cite{kn:ong}, to which we refer the reader for the details. To summarize:

\medskip

$\bullet$ If we fix the Lorentzian horizon at $r_h=1$ and repeat the calculation above, we find that the Lorentzian action is \emph{negative} (for sufficiently large electric charge) in some range of $r$ if $0 < \alpha < \alpha_c < 1$, where $\alpha_c$ is a critical value\footnote{One must be careful when normalizing the position of the horizon $r_h$ (and $r_h^E$), since although this preserves the qualitative behavior of the brane action, quantitative features can be affected. This means that given a normalization value of $r_h$, say $r_h=s > 0$, the critical value of $\alpha$ is actually dependent on $s$. The actual critical value should be defined as the smallest value of $\alpha$ such that the brane action is non-negative (for all admissible values of $b$), independent of $s$. The asymptotic behavior of the action given by equation (\ref{OYC}) remains unaffected by normalization choice; more specifically, only $\text{const}(\alpha)$ is normalization-dependent.}
of $\alpha$ (which we estimate numerically at around 0.53) although it eventually turns around and asymptotically grows linearly in $r$ according to the same expression above (equation (\ref{OYC})), which only depends on the \emph{square} of $b$. (In five spacetime dimensions, the action grows logarithmically in $r$ \cite{kn:ong}.) Of course, $\text{const}(\alpha)$ is different in the Lorentzian case.
This is quite different to the case without a dilaton, in which, for sufficiently large charge, the action becomes negative and \emph{stays} negative, that is, $\text{const}(\alpha)$ is negative if $\alpha=0$; see Figure (\ref{fig2}). As was pointed out by Maldacena and Maoz \cite{kn:maldacena}, actions of this kind represent a relatively benign form of instability, since the region with negative action is \emph{finite}: presumably the system evolves to some nearby state rather than getting entirely out of control. Nevertheless, this does suggest that, at high values of the electric charge, values of $\alpha$ smaller than $\alpha_c$ should not be considered internally consistent in the sense we are studying here. In short, the situation here has a similar physical interpretation to the one studied in the preceding section: when $\alpha < \alpha_c$, holographic consistency imposes an upper bound on $\mu_B/T$, the ratio of the baryonic chemical potential to the temperature. (This bound will take the form of an $\alpha$-dependent version of the inequality given by (\ref{O}) above.)

\medskip

$\bullet$ If $\alpha > \alpha_c$, then the Lorentzian action is positive for all values of the charge: holographic consistency imposes no restrictions in either the Euclidean or the Lorentzian case.

\medskip

\begin{figure}[!h]
\centering
\includegraphics[width=6.0in]{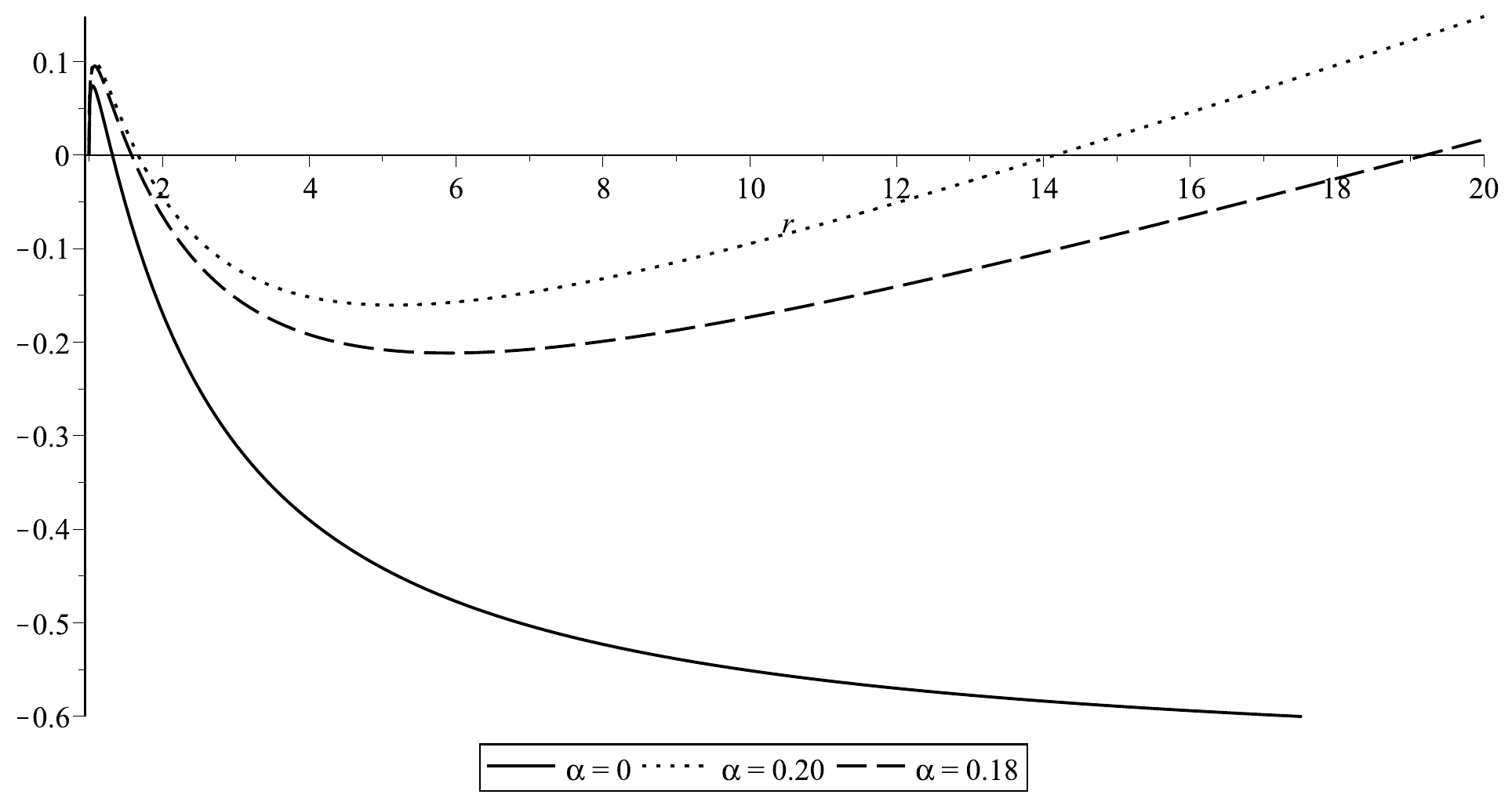}
\caption{\label{fig2}
The corresponding Lorentzian brane actions $\mathfrak{S}^\text{L}(r)$ of the Gao-Zhang black hole (up to a positive factor) with various values of the dilaton coupling parameter $\alpha$, the same charge parameter $b=0.7$, and horizons held fixed at $r_h=1$. Here we set $L=1$, however the value of $L$ only contributes an overall factor to the action, and thus does not affect its sign. Note that the case $\alpha=0$ reduces to an electrically charged Reissner-Nordstr\"om black hole, discussed in subsection (2.2), which in this case tends asymptotically to the value $-0.6$. For $\alpha \neq 0$, the action always grows asymptotically in $r$.
}
\end{figure}

These results are potentially of great interest in the application of these black holes to the question of the thermalization of the quark-gluon plasma, as studied in \cite{kn:zhang}. There it was found\footnote{It is true that the metric used in \cite{kn:zhang} is not the Gao-Zhang metric itself, but rather a Vaidya-like deformation of it. However, we doubt that this will change the qualitative conclusions we are drawing here.} that there is an ($\alpha$-dependent) upper bound on $\mu_B/T$ when $\alpha > 1$, but no restriction whatever when $\alpha < 1$; this bound is not due to any instability, but rather simply to the form taken by $\mu_B/T$ as a function of another parameter (the saturation time; see Figure 3 in \cite{kn:zhang}). In other words, there is a bound on $\mu_B/T$ when the dilaton is strongly coupled ($\alpha > 1$). What we are finding here is that there is \emph{also} such a bound, imposed by holographic consistency, in the weak dilaton coupling regime ($0 < \alpha < \alpha_c < 1$).

The holographic bound, in the weak-coupling case, is presumably weaker (that is, higher) than in the case considered in the preceding section. If forthcoming data should violate the bound discussed above (inequality (\ref{O})), then one might try to use the $\alpha$-dependent version of it to avoid the conflict. The role of holography would then be to put a lower bound on $\alpha$. We conjecture that the strong-coupling bound might likewise be used to put a useful \emph{upper} bound on it.  The task then would be to use the range of $\alpha$ values so obtained to constrain the values of parameters more directly related to observations, such as thermalization times. This has yet to be done.

In summary, the situation in this case is less clear, since the theory is less fully developed than in our earlier examples; all we can say definitely is that, while Euclidean consistency is certainly satisfied here, Lorentzian consistency is not automatic and may ultimately prove useful in constraining the key parameter $\alpha$. One can hope that, when the subject of holographic thermalization (or ``dilatonic holography'' more generally) is more mature, it will be possible to investigate more fully whether the Lorentzian consistency condition is satisfied here. At present, there is no reason to suspect otherwise.

\subsection*{{\textsf{4. The Einstein Case}}}
We saw in the preceding sections that, because (some) black hole parameters are affected by complexification, good behaviour in the Euclidean case does not necessarily ensure equally good behaviour in the Lorentzian case. One might be tempted to argue that this problem is due to the presence of matter in the bulk, since, after all, the difficulty arises from the presence of electromagnetic fields, and from the complexification of the electric charge. Unfortunately that is not so: the various bulk spacetimes endowed with \emph{angular momentum} are still Einstein manifolds in some cases, but, in every case, the angular momentum parameter has to be complexified in the passage to the Lorentzian domain, and we will see that this can have consequences similar to those associated with complexifying electric charge.

We will consider two cases: topologically spherical event horizons, and their planar counterparts.
\subsubsection*{{\textsf{4.1 AdS-Dyonic-Kerr-Newman with Topologically Spherical Event Horizon }}}
The four-dimensional asymptotically AdS dyonic Kerr-Newman metric with a topologically spherical event horizon (which we continue to indicate by a ($+1$) superscript, though the actual geometry is not that of a round sphere) \cite{kn:carter} takes the form, in Boyer-Lindquist-like coordinates,
\begin{flalign}\label{P}
g(\m{AdSdyKN^{+1}_{4})} = &- {\Delta_r \over \rho^2}\Bigg[\,\m{d}t \; - \; {a \over \Xi}\m{sin}^2\theta \,\m{d}\phi\Bigg]^2\;+\;{\rho^2 \over \Delta_r}\m{d}r^2\;+\;{\rho^2 \over \Delta_{\theta}}\m{d}\theta^2 \\ \notag \,\,\,\,&+\;{\m{sin}^2\theta \,\Delta_{\theta} \over \rho^2}\Bigg[a\,\m{d}t \; - \;{r^2\,+\,a^2 \over \Xi}\,\m{d}\phi\Bigg]^2,
\end{flalign}
where again the ``dy'' denotes ``dyonic'' and where
\begin{eqnarray}\label{eq:Q}
\rho^2& = & r^2\;+\;a^2\m{cos}^2\theta, \nonumber\\
\Delta_r & = & (r^2+a^2)\Big(1 + {r^2\over L^2}\Big) - 2Mr + {Q^2 + P^2\over 4\pi},\nonumber\\
\Delta_{\theta}& = & 1 - {a^2\over L^2} \, \m{cos}^2\theta, \nonumber\\
\Xi & = & 1 - {a^2\over L^2}.
\end{eqnarray}
Here $- 1/L^2$ is the asymptotic curvature, $a$ is the angular momentum/mass ratio, and $M, Q$, and $P$ are related to the physical mass $E$, electric charge $q$, and magnetic charge $p$, by (see \cite{kn:gibperry})
\begin{equation}\label{R}
E\;=\;M/\Xi^2, \;\;\;\;\;q\;=\;Q/\Xi,\;\;\;\;\;p\;=P/\Xi ;
\end{equation}
note that all of these depend on the angular momentum. As before, this metric is not, in general, an Einstein metric; but it is Einstein when $P = Q = 0$, for \emph{any} value of $a$. That is the case in which we are most interested here; but it will be interesting to retain $Q$ and $P$ so as to study the general case.

This black hole corresponds holographically to a \emph{rotating} quark-gluon plasma \cite{kn:sonner,kn:schalm}. In fact, it is expected that, under some circumstances (connected with the viscosity of the plasma), the plasma produced in a peripheral heavy-ion collision will indeed have a strong rotational motion \cite{kn:KelvinHelm,kn:viscous}, so this geometry supplies a holographic description of that motion. (In other cases, the internal motion of the plasma is a shearing rather than a rotation: see the next section.)

Now the geometry of the spacetime described by (\ref{P}) is, unless we impose a certain condition, rather peculiar. In particular, consider the function $\Delta_{\theta}$: in general, this function does not have a fixed sign, being positive in directions near the equator, but possibly negative towards the poles. If indeed $\Delta_{\theta}$ does change sign in this way, then the signature of the metric (outside the event horizon) changes from $(-\,+\,+\,+)$ to $(-\,+\,-\,-)$ as one rotates from the equator to the poles, so that, in particular, the geometry at conformal infinity ($r\,\rightarrow \, \infty$) has signature $(-\,-\,-)$ in some directions, $(-\,+\,+)$ in others. This bizarre behaviour is unphysical from a holographic point of view, indeed probably from any point of view\footnote{Note that this is not like the more familiar signature change discussed in, for example, \cite{kn:ellis}, or more recently in \cite{kn:bojo}.}, so we have to impose the condition\footnote{The case with $a^2/L^2 = 1$ is excluded because then, by the equations (\ref{R}), the parameters $M, Q, P$ have no physical interpretation.}
\begin{equation}\label{S}
a^2/L^2 \;<\; 1;
\end{equation}
this strange relation between the angular momentum/mass ratio of the black hole and the asymptotic spacetime curvature is the only way to ensure that $\Delta_{\theta}$ remains positive for all $\theta$. This apparently recondite point will in fact be crucial for our later discussion.

The electromagnetic potential form in the exterior spacetime is given by (see \cite{kn:frolov} for the asymptotically flat case)
\begin{equation}\label{T}
A = -\,{Q\,\Xi\, r\over 4\pi\rho^2}\left[\m{d}t-{a\,\m{sin}^2\theta\over \Xi}\m{d}\phi\right]-{P\,\Xi\,\m{cos} \theta\over 4\pi\rho^2} \left[a\,\m{d}t - {r^2+a^2 \over \Xi}\m{d}\phi\right].
\end{equation}
From this one sees that $a$ must be complexified ($a\,\rightarrow \, -ia$) along with $Q$ ($Q\,\rightarrow \, -iQ$) when passing to the Euclidean version ($t\,\rightarrow \, it$), while, as usual, $P$ must not.

Up to the usual overall positive factor, $\mathfrak{S^{\m{E}}}(r)$ for this geometry takes the form \cite{kn:74}
\begin{eqnarray}\label{U}
\mathfrak{S^{\m{E}}}(\m{AdSdyKN^{0}_{4})}(r) & = & \left\{r\sqrt{(r^2-a^2)\Big(1 + {r^2\over L^2}\Big) - 2Mr - Q^2 +P^2}\right.\,\,\times  \nonumber\\ &&\;\;\;\left.\Bigg[\sqrt{1-{a^2\over r^2}}+ {r\over a}\, \m{arcsin}{a\over r}\Bigg]\right\}-\;{2r^3\over L}\Bigg[1 - {a^2\over r^2}\Bigg] \nonumber \\ && \;\;\;+\;{2(r_h^E)^3\over L}\Bigg[1 - {a^2\over (r_h^E)^2}\Bigg],
\end{eqnarray}
where $r_h^E$ has the usual meaning.

Extensive numerical tests strongly suggest that this is a positive function of $r$ for all $r > r_h^E$. One can see that this is the case at large $r$ by expressing this function in the form
\begin{equation}\label{eq:V}
\mathfrak{S^{\m{E}}}(\m{AdSdyKN^{0}_{4}})(r) \;=\; rL \left(1 + {2a^2 \over 3L^2}\Bigg)\;+\;{2(r_h^E)^3\over L}\Bigg(1 - {a^2\over (r_h^E)^2}\right) - 2ML\;+\;O(1/r).
\end{equation}
One sees that there are two terms that do not decay towards infinity: a linear term and a constant term. The dominant term here is of course the one linear in $r$, and it is clearly positive, so the function is certainly positive at large $r$; in fact it is almost certainly positive everywhere, so the consistency condition, equation (\ref{B}), is satisfied. That had to be so when $Q = P = 0$, since $g^E(\m{AdSdyKN^{+1}_{4})}$, the Euclidean version of the metric here, is an Einstein metric in that case, and it induces a conformal structure at infinity which evidently has a positive Yamabe invariant: so Wang's theorem applies. However, it was not clear that it would hold in the charged case.

More remarkable, because (as we have seen) it does not follow from the result in the Euclidean case, is that the Lorentzian version of this quantity,
\begin{equation}\label{eq:W}
\mathfrak{S^{\m{L}}}(\m{AdSdyKN^{0}_{4}})(r) \;=\; rL \left(1 - {2a^2 \over 3L^2}\right)\;+\;{2(r_h)^3\over L}\left(1 + {a^2\over (r_h)^2}\right) - 2ML\;+\;O(1/r),
\end{equation}
is \emph{also} positive at large values of $r$; in fact, again, the numerical evidence \cite{kn:74} very strongly suggests that it is positive everywhere outside the black hole. This would not be so if it were possible for the angular momentum/mass ratio $a$ to satisfy $a^2/L^2 > 3/2,$ but that is forbidden by the inequality (\ref{S}) given above. Notice that the complexification of $Q$ plays no role here: it does of course affect the numerical details (because reversing the sign of $Q^2$ affects the value of $r$ at the event horizon, that is, $r_h \neq r_h^E$) but it does not affect the sign of the dominant term. For these black holes, then, it does not matter whether the spacetime is Einstein or not.

Thus we see that this system respects holographic consistency, for all (physical) values of the parameters, in both the Euclidean and Lorentzian versions of the geometry, even in the non-Einstein case. It is striking, however, that in the Lorentzian case we had a narrow escape: the situation is saved only by the technical condition that conformal infinity should have a consistent signature, expressed by the inequality (\ref{S}). Again we see that the Lorentzian case is more delicate than its Euclidean counterpart.

\subsubsection*{{\textsf{4.2 Dyonic KMV$_4^0$ with Planar or Toral Event Horizon }}}
The dyonic planar AdS black hole metric discussed earlier (equation (\ref{E})) can be endowed with angular momentum; in fact, this can be done in many ways: see
\cite{kn:chrusc,0401081, 0901.2574, 0904.1566, 1107.3677, 1201.3098} for detailed discussions of the mathematical and physical ramifications of this. However, if we focus on the most physically interesting case, in which the boundary is conformally flat, then \cite{kn:shear,kn:76} the possibilities are enormously restricted. In essence, there are two possible families. The first was obtained in the zero-charge case by Klemm, Moretti, and Vanzo \cite{kn:klemm}; with the addition of electric and magnetic charges, we call these the ``dyonic KMV$_4^0$'' or ``dyKMV$_4^0$'' metrics:
\begin{equation}\label{X}
g(\m{dyKMV_4^0}) = - {\Delta_r\Delta_{\psi}\rho^2\over \Sigma^2}\,\m{d}t^2\;+\;{\rho^2 \over \Delta_r}\m{d}r^2\;+\;{\rho^2 \over \Delta_{\psi}}\m{d}\psi^2 \;+\;{\Sigma^2 \over \rho^2}\left[\omega\,\m{d}t \; - \;\m{d}\zeta\right]^2,
\end{equation}
where the coordinates and parameters are as in equation (\ref{E}) (with the addition of $a$, the angular momentum/mass ratio), and where
\begin{eqnarray}\label{Y}
\rho^2& = & r^2\;+\;a^2\psi^2, \nonumber\\
\Delta_r & = & a^2+ {r^4\over L^2} - 8\pi M^* r + 4\pi (Q^{*2}+P^{*2}),\nonumber\\
\Delta_{\psi}& = & 1 +{a^2 \psi^4\over L^2},\nonumber\\
\Sigma^2 & = & r^4\Delta_{\psi} - a^2\psi^4\Delta_r,\nonumber\\
\omega & = & {\Delta_r\psi^2\,+\,r^2\Delta_{\psi}\over \Sigma^2}\,a.
\end{eqnarray}
As in the preceding section, this is an Einstein metric for any value of $a$, provided that $Q^* = P^* = 0$.

The second family of metrics with angular momentum and with conformally flat boundaries is obtained by adding a parameter similar in some ways to NUT charge: these are the ``$\ell$dyKMV$_4^0$'' metrics introduced (without magnetic charge) in \cite{kn:shear}. As they are rather more complicated than the dyKMV$_4^0$ metrics, and as they do not lead to different conclusions, we shall not discuss them here; see below.

The electromagnetic potential form in the dyKMV$_4^0$ case is
\begin{equation}\label{Z}
A =-\, {1\over \rho^2L}\left[(Q^*r + aP^*\psi)\m{d} t \;+\;(aQ^*r\psi^2 - P^*\psi r^2)\m{d}\zeta\right],
\end{equation}
from which we see that the usual pattern of complexifications continues to hold here.

These black holes have a very remarkable property: like any black hole with angular momentum, they induce frame dragging effects in the surrounding spacetime, but here the frame-dragging effect persists to conformal infinity; yet it is not a uniform rotation there, as it is in the topologically spherical case considered above. Instead, the frame-dragging mimics a \emph{shearing} motion. Under some circumstances (related, as before, to the viscosity of the plasma), the plasma produced by a peripheral heavy-ion collision does indeed take the form of a shearing motion \cite{kn:liang,kn:bec,kn:huang} (see \cite{kn:csernai,1405.7283,1406.1017,1406.1153,1503.03247} for more recent developments). Thus one can use the KMV metrics and their generalizations to give a holographic account of the internal motion of the plasma in these situations \cite{kn:77,kn:shear}.

The dimensionless velocity of the shearing plasma described by the dyKMV$_4^0$ metric is given by
\begin{equation}\label{ALPHA}
v(x) \;=\; a\psi^2/L;
\end{equation}
this corresponds to a motion within the plasma, increasing away from the $\psi = 0$ axis, which corresponds to the axis of the collision in the dual system. Causality therefore imposes the bound
\begin{equation}\label{BETA}
\psi\;<\;\Psi \; \equiv \; \sqrt{L/a}.
\end{equation}
Note carefully that $\Psi$ is just a special numerical value of the spacelike coordinate $\psi$, which of course is never complexified; so we must \emph{not} complexify $a$ in this formula when we pass to the Euclidean geometry.

When we do move to the Euclidean case, we find again that $\psi$ must still satisfy the inequality (\ref{BETA}), since otherwise various pathologies will arise: for example, if (\ref{BETA}) is not enforced, then the Euclidean version of $\rho^2$ can be negative at some values of $r$, and the Euclidean version of  $\Delta_{\psi}$ (given by $1 - (a^2 \psi^4/L^2)$) is negative for some values of $\psi$; so that, in particular, the coefficients of d$r^2$ and d$\psi^2$ in the ``Euclidean'' version of the metric will in that case have opposite signs, which is a contradiction.

Thus in both cases $\psi$ ranges between 0 and $\Psi$, so areas and volumes can now be evaluated accordingly: one then finds that the Euclidean quantity $\mathfrak{S^{\m{E}}}(r)$ in this case takes the form
\begin{eqnarray}\label{GAMMA}
\mathfrak{S^{\m{E}}}(\m{dyKMV_0})(r) & = & \left\{{r^2\over 2} \sqrt{-a^2+ {r^4\over L^2} - 8\pi M^*r+ 4\pi (-\,Q^{*2}+P^{*2})}\right. \times \nonumber \\
&&
\;\;\;\;
\left.\left[{1 \over a}\m{arcsin}\Big({a\Psi\over r}\Big) + {\Psi \over r}\sqrt{1-{a^2\Psi^2\over r^2}}\right]\right\} \nonumber \\
&&
\;\;\;\;
-  {1\over L}\left[\Psi(r^3 - (r_h^E)^3) - a^2\Psi^3(r-r_h^E)\right],
\end{eqnarray}
where $r_h^E$ locates the Euclidean ``event horizon''. As in the preceding section, numerical evidence strongly suggests that this function is positive everywhere beyond the Euclidean event horizon; one can see this directly at large $r$:
\begin{equation}\label{DELTA}
\mathfrak{S^{\m{E}}}(\m{dyKMV_0})(r) \;=\; {5a^2\Psi^3 \over 6L}\,r \;+\;{\Psi r^E_{h} \over L}\left[\,(r^E_{h})^2 \;-\;a^2\Psi^2 \;-\;{4\pi M^*L^2 \over r^E_{h}}\right] \;+\; O(1/r) .
\end{equation}
As in the topologically spherical case, there are two terms that do not decay towards infinity, the linear term being the dominant one; and clearly the function is positive at large $r$ for all values of $a$ and of the charges. But when we turn to the Lorentzian version, we find a result very different from the topologically spherical case: we have
\begin{equation}\label{EPSILON}
\mathfrak{S^{\m{L}}}(\m{dyKMV_0})(r) \;=\; { -\,5a^2\Psi^3 \over 6L}\,r \;+\;{\Psi r_{h} \over L}\left[\,r_{h}^2 \;+\;a^2\Psi^2 \;-\;{4\pi M^*L^2 \over r_{h}}\right] \;+\; O(1/r) .
\end{equation}
$\mathfrak{S^{\m{L}}}(\m{dyKMV_0})(r)$ is positive near to the event horizon of the black hole, but this expression shows that it is negative (in fact, unbounded below) far from it. The Lorentzian system is \emph{unstable} for all non-zero values of all parameters. (The situation for the other family of metrics mentioned earlier, the $\ell$dyKMV$_4^0$ metrics, is essentially the same.)

In particular, if we focus on the $Q^* = P^* = 0$ case, we have here an example of an Einstein metric in the bulk which (in the Euclidean case, after compactification) has non-negative Yamabe invariant at infinity, so, by Wang's theorem, the Euclidean version of the system had to be consistent; but the Lorentzian version nevertheless misbehaves, for all values of the angular momentum, if the corresponding plasma is sufficiently long-lived.

In fact, however, the relevant plasma here, one which is endowed with a very large angular momentum density, is not the cosmic plasma we considered in section 2 of this work; instead it is the plasma produced in a heavy ion collision. Such plasmas only survive for a very short time, a few femtometres/c, so it is not clear that the Lorentzian instability has sufficient time to manifest itself. In fact, a plasma with a violent internal motion might well be subject to hydrodynamic instabilities analogous to or generalizing the well-known \emph{Kelvin--Helmholtz instability} \cite{kn:KelvinHelm}; and it may be that such instabilities do in fact set in as the plasma hadronizes. Thus, we should interpret Lorentzian holographic consistency as an upper bound on the \emph{time} during which a hydrodynamic model of the plasma is valid. In order to judge whether consistency is violated here, one would need to estimate the time required for the instability to be established. A holographic method of doing so was proposed in \cite{kn:77} (see also \cite{kn:shear}), and in fact preliminary estimates do suggest that the instability time scale is approximately the same as that of hadronization.

Again, therefore, we conclude that, while Lorentzian consistency is not (unlike Euclidean consistency) guaranteed in this case, \emph{in practice} it does not fail ---$\,$ though it very easily might have done so.

\subsection*{{\textsf{5. Conclusion: Consistency as a Law of Physics}}}
It has long been hoped that at least some of the laws of physics might be found to follow inevitably from the requirements of internal mathematical consistency in some unified theory. We propose that Ferrari's Euclidean consistency condition (\ref{A}) should be considered in this manner. We have seen that doing so, and making the natural move of imposing the analogous Lorentzian condition, constrains a wide variety of quark-gluon plasmas (in particular, the quite different plasmas occurring in the early Universe and in heavy ion collisions) in very remarkable ways. One is struck particularly by the fact that observable systems repeatedly come close to violating these constraints, without ever actually doing so.

In the title of this work, we asked a question: ``When is Holography Consistent?''. The answer appears to be, ``Always, at least in all of the various examples we have considered.'' It seems that holographic consistency, in the specific form of the ``isoperimetric inequality'' (\ref{B}), has the character of a law of physics. It will be interesting to investigate whether it continues to hold as more data accumulate and in other applications.

\addtocounter{section}{1}
\section*{\large{\textsf{Acknowledgement}}}
BMc is grateful to Dr. Soon Wanmei, J.L. McInnes, and C.Y. McInnes, for helpful discussions. YCO thanks Nordita for supporting his travel to Singapore, where part of this work was completed.

\end{document}